\documentclass[prl,preprint,showpacs,amsmath,amssymb]{revtex4} 

\usepackage{graphicx}
\begin{document}

\title{Infuence   of  correlation  effects on the  of
magneto-optical   properties   of   half-metallic  ferromagnet   NiMnSb}
\author{S.~Chadov} \author{J.~Min\'{a}r} \author{H.~Ebert}
\affiliation{Dept.~Chemie    und    Biochemie,
 Physikalische  Chemie,  Universit\"at  M\"unchen,  Butenandtstr.   5-13,
D-81377 M\"unchen,  Germany} \author{A.~Perlov} \affiliation{Accelrys,
334 Cambridge Science Park, CB4 0WN Cambridge, England}
\author{L.~Chioncel $^\dagger$} \affiliation{Institute
of  Theoretical  Physics,  Technical  University of  Graz,  A-8010 Graz,
Austria} \author{M.I.~Katsnelson} \affiliation{University  of Nijmegen,
NL-6525   ED  Nijmegen,   The   Netherlands}  \author{A.I.~Lichtenstein}
\affiliation{Institute  of Theoretical  Physics, University  of Hamburg,
Germany}

\begin{abstract}  The magneto-optical spectra of  NiMnSb were  calculated in
the framework  of the Local  Spin Density Approximation  (LSDA) combined
with Dynamical Mean-Field Theory (DMFT).  Comparing with results based
on plain LSDA, an additional  account of many-body  correlations via
DMFT results  in a noticably improved agreement  of the theoretical
Kerr-rotation and ellipticity spectra with corresponding experimental data.
\end{abstract}


\maketitle

Since the  discovery of the giant magneto-optical
Kerr  effect (MOKE) in  PtMnSb \cite{EBJ+83}  magneto-optical properties
became  an important  issue for the Mn-based  family of  Heusler alloys
\cite{Eng83,OKM87,EMW+94,EM97,KC95,OAK+95,AOY+97,GWK+99,PCF06}.
However, despite  of the similar  structure, the group  of isoelectronic
alloys  PtMnSb, NiMnSb and  PdMnSb show  quite different maximum 
amplitudes in their MOKE spectra \cite{EBJ+83,Eng83}. A theoretical description  of the observed
difference of the MOKE spectra became possible within {\em ab-initio} band-structure
calculations  \cite{KUH94,UKK95,AOY+97,PCF06}. However, although  the  various
calculated  MOKE  spectra  give  reasonable qualitative  agreement  with
 experiment,   one  can   notice  that   there  exist   several  systematic
discrepancies generally ascribed to the use of local spin density
approximation (LSDA). In particular,
for NiMnSb,  one can see  that the low-energy  peak of the Kerr rotation
spectra at 1.4~eV is shifted to a regime of 
1.6-2~eV. Also  there is a noticable  deviation of the  amplitude for
the peak at 4~eV as well as in the intermediate energy regime.
Among  other  reasons, the  discrepancies encountered  in LSDA-based
results  could   appear  due  to   an  insufficient  treatment   of  electronic
correlations.    For   example,   as   it   was   shown   for   bcc   Ni
\cite{PCE03,PCE+04},  the  account of  local  dynamical correlations  is
essentially important for a proper description of its MOKE spectra.
In the case of NiMnSb the main contribution to the optical transitions comes
from the $d$-shell of Mn which supplies the unoccupied part of the
density of states (DOS). At the
same  time $d$-electrons of  Mn should  be treated  as locally-correlated
\cite{CKG+03}. Based on this  supposition one can expect an improvement
of  MO  spectra  in NiMnSb  by an  appropriate account  of  local
correlation effects for the Mn  $d$-shell in band-structure calculations.  In
the present work  the latter is  implemented within the  so-called Dynamical
Mean Field Theory approach (DMFT) \cite{GKK+96KV04}.

In  our calculations the  central quantity  is the  optical conductivity
tensor  $\sigma_{\lambda\lambda'}$, as all optical  and magneto-optical
properties,  can  be   expressed   through  its cartesian components ($\rm\lambda=\{x,y,z\}$). In
particular for the complex Kerr angle, that combines Kerr rotation
$\theta_{\rm  K}$ and ellipticity $\varepsilon_{\rm K}$, the
following expression can be used \cite{ES75}:
\begin{equation} \rm \theta_K (\omega ) + {\rm i}\,\varepsilon_K (\omega
)  \approx  \frac{   -   \sigma_{xy}  (\omega   )   }{\sigma_{xx}(\omega   )
\sqrt{\displaystyle 1 +  \frac{4 \pi \rm i}{\omega }\,\sigma_{xx}(\omega
)}} \, .
\label{defkerr}
\end{equation} In order to calculate the optical conductivity we use the
expression for the hermitian  component derived for the zero-temperature
case \cite{SW99} by implementing  the Green's function formalism in Kubo's
linear response theory \cite{Kub57Gre58}:
\begin{eqnarray}  \sigma_{\lambda\lambda'}^{(1)}(\omega)  =-\frac{1}{\pi
V}\!\!\!\int\limits_{\epsilon_{\rm         F}-\hbar\omega}^{\epsilon_{\rm
F}}\!\!\!\!\!{\rm    d}\epsilon~{\rm   Tr}\left\{   \hat{J}^\lambda\,\Im
G(\epsilon+\hbar\omega)\,\hat{J}^{\lambda'}\,\Im G(\epsilon)\right\}\!,~
  \label{kubo_GF}
\end{eqnarray}  where  $V$  is  the  volume of  the  spatial  averaging,
$\hat{J}^{\lambda}$     is    the    current-density    operator. $\Im G(\epsilon)$  stands for
the   antihermitian   part   of   the  retarded   one-electron   Green's
function.   The   hermitian   part   of  the   conductivity   tensor
$\sigma_{\lambda\lambda'}^{(2)}(\omega)$   can  be   obtained   via  the
Kramers-Kronig relations.

Formulation (\ref{kubo_GF}) is used because it allows to include
straightforwardly all correlations in the one-particle Green's function via
the Dyson equation:
\begin{eqnarray}
\left[\epsilon-\hat{H_0}-\hat{\Sigma}(\epsilon)\right]\hat{G}(\epsilon)=\hat{I}\,,
\label{defgf}
\end{eqnarray} where  $\hat{H_0}$ is LSDA-based one-particle  Hamiltonian including
the kinetic energy, electron-ion  interaction and the Hartree potential,
while the  self-energy operator $\hat{\Sigma}$ describes  all static and
dynamic effects of electron-electron exchange and correlations. The most
popular  approximation  nowadays  for  the  self-energy  is  DMFT  which
introduces it   as  a  local,   energy-independent  exchange-correlation
potential  $V_{\rm xc}(r)$. As  the introduction  of such  an additional
potential does  not change the  properties of $H_0$ we  will incorporate
this  potential  to $H_{\rm  LSDA}$  and  subtract  this term  from  the
self-energy  operator.    This  means  that   the  self-energy  $\Sigma$
describes  exchange and  correlation  effects not  accounted for  within
LSDA.

The most straightforward and accurate way to solve Eq.\,(\ref{defgf}) is
to  use  the  Korringa-Kohn-Rostoker  Green's function  method  (KKR-GF)
\cite{HZE+98}.  However subsequent calculations  of  optical properties
within   the KKR-GF   method  become   very   time-consuming   due  to   the
energy-dependent  matrix  elements   of  the current-density  operator.
A possibile alternative is to use  the so-called variational,  or fixed
basis-set  methods.  Within  such an  approach the  Green's  function is
represented   as  a   sum  over   energy  independent   basis  functions
$|i\rangle$:
\begin{eqnarray}                  G(\epsilon)&=&\sum_{ij}|i\rangle\langle
i|\frac{1}{\epsilon-H_{\rm  LSDA}-\Sigma(\epsilon)}|  j\rangle\langle j|
\\ &=& \sum_{ij}|i\rangle G_{ij}(\epsilon)\langle j|\,,\nonumber
\label{gfmat}
\end{eqnarray} with the Green's function matrix defined as
\begin{eqnarray}       G_{ij}(\epsilon)      =       \left[      \langle
i|j\rangle\epsilon-\langle    i|\hat{H}_{\rm   LSDA}|j\rangle   -\langle
i|\hat{\Sigma}(\epsilon)|j\rangle \right]^{-1}\,.
\end{eqnarray} The dynamical  correlation correction to the one-particle
hamiltonian  $H_{\rm  LSDA}$  is  represented via  the  energy-dependent
self-energy  operator $\hat{\Sigma}({\vec  r},  {\vec r\,'},  \epsilon)$
which in general is a non-local quantity. Here, the  self-energy is  calculated  via  the DMFT  approach
 \cite{GKK+96KV04} that accounts only  for local (on-site)
 correlations which  are described  within the Anderson impurity model (AIM)
 \cite{And61}.  By linking the Green's function of the  effective
 impurity to  the single-site Green's  function derived from the $\bf
 k$-space  summation DMFT provides the self-consistent method to
 determine  the  self-energy  accounting  for  on-site
 correlations. The actual approximation made in DMFT is the substitution
 of the  non-local ($\bf k$-dependent) self-energy  by the single 
on-site  component.  The latter
corresponds to  the limit of  the infinite coordination.
Fortunately,  in  many  cases for 3-dimensional  systems this is rather  good
approximation \cite{MV89}.

Although  the formalism  to  be  presented below  is  primarily used  in
connection with a site-diagonal  self-energy, it should be stressed that
any more complex self-energy can be used  as well. In particular the formalism
is  able to  deal with  a  site-non-diagonal self-energy  occuring  for example
within the GW approach \cite{BAG03}.

Dealing with crystals one can  make use of Bloch's theorem when choosing
basic  functions   $|i_{\bf  k}\rangle$.   This  leads  to   the  $\bf
k$-dependent Green's function matrix:
\begin{eqnarray}    G^{\bf    k}_{ij}(\epsilon)    =   \left[    \langle
i_{\bf_k}|j_{\bf_k}\rangle\epsilon-\langle         i_{\bf_k}|\hat{H}_{\rm
LSDA}|j_{\bf_k}\rangle                                           -\langle
i_{\bf_k}|\hat{\Sigma}(\epsilon)|j_{\bf_k}\rangle\right]^{-1}\!\!\!\!\!\!.
\label{gfmatk}
\end{eqnarray}
The efficiency and accuracy of  the approach is determined by the choice
of  $|i_{\bf  k}\rangle$.  One of  the  most computationally  efficient
variational methods is  the LMTO \cite{And75} which allows  one to get a
rather accurate description of the  valence and conduction bands in the range
of  about 10~eV, which  is  enough for  the calculation  of the  optical
spectra ($\hbar\omega<\,$6-8~eV)  . As  the method introduces  site- and
angular momentum -dependent basis functions
\begin{eqnarray}   \chi^{\bf   k}_{lm\bf   R}(\vec  r)   &=&   \Phi^{\rm
h}_{lm}(\vec   r-{\bf  R})\nonumber   \\  &+&\sum_{lm',\bf   R'}  h^{\bf
k}_{lm{\bf R}, lm'{\bf R'}}\Phi^{\rm t}_{lm'}(\vec r-{\bf R'})\,,
\label{lmtobasis}
\end{eqnarray} it perfectly fits to any single-site approximation of the
self-energy.   The superscripts  "h"  and "t"  stand  for the  so-called
"head" and "tail" parts of the basis functions.

As the $\langle i |j\rangle$  and $\langle i | \hat{H} |j\rangle$ matrix
elements in  Eq.\,(\ref{gfmatk}) are energy-independent it  is enough to
calculate them only once for each $\bf k$-point.

In the framework of the DMFT the self-energy operator can be expressed in
the form:
\begin{eqnarray} \hat{\Sigma}(\vec r, \vec r\,',\epsilon)&=&\sum_{l, \bf
RR'}\delta_{\bf          RR'}\sum_{mm'}\Theta(\left|\vec          r-{\bf
R}\right|)\,Y^{*}_{lm}(\vec  r-{\bf  R})\nonumber \\  \times~\Sigma^{\bf
RR'}_{lmm'}&&\!\!\!\!\!\!\!\!\!\!\!(\vec       r,       \vec       r\,',
\epsilon)~\Theta(\left|\vec  r\,'-{\bf R'}\right|)\,Y_{lm'}(\vec r'-{\bf
R'})\,,
\label{dmftlmtosigma}
\end{eqnarray}  where $\Theta(r)=1$  if $\vec  r$ is  inside  the atomic
sphere  and zero otherwise.   Due to  the special  choice of  LMTO basis
functions  (\ref{lmtobasis}) only  the  "head" component  will give  the
significant  contribution  to the  matrix  elements  of the  self-energy
operator  (\ref{dmftlmtosigma})  leading  to  an extremely  simple  $\bf
k$-independent expression:
\begin{eqnarray}               \langle               \chi^{\bf_k}_{lm\bf
R}|\hat{\Sigma}|\chi^{\bf_k}_{lm'\bf
R}\rangle&\approx&\int{d^3}r{d^3}r'\nonumber   \\   \times~\Phi^{\rm   h
*}_{lm}(\vec r-{\bf R})&&\!\!\!\!\!\!\!\!\!\!\Sigma^{\bf RR}_{lmm'}(\vec
r, \vec r\,', \epsilon)~\Phi^{\rm h}_{lm'}(\vec r-{\bf R})\,.
\end{eqnarray}  
The   accuracy  test  of  this  approach   was  done  in
\cite{PCE03}  and  the  error  was  found  to be  within  5\%  which  is
substantially less  than the approximations  made for the  estimation of
the self-energy itself.

The  self-energy is  calculated  by using as an AIM-solver the  so-called
Spin-Polarised T-matrix  plus Fluctuation Exchange  (SPTF) approximation
\cite{BS89,LK98KL99KL02}.   SPTF  is   a  perturbative  appoach  which
provides an analytical technique to sum the infinite set of the Feynmann
diagramms for the  several types of interactions in  the uniform electron
gas.  Fortunately these sets of diagrams very often appear to be
sufficient to describe  the effects caused by dynamical  correlations in 
moderately-correlated shells  like $3d$-electrons in  transition metals
\cite{PCE+04,MCP+05} as well as  in systems with strong correlations
\cite{PKL05}.  This  makes  it  a  very attractive  alternative  to  the
so-called  Quantum Monte-Carlo  (QMC) technique  \cite{HF86HNB+01} that
sums all  possible sets of  diagramms of perturbation theory  and
therefore is much more time consuming.

The DMFT scheme is implemented within the KKR-GF method which allows 
to use the  advantages of  the  scattering theory  formulation  in the  Green's
function construction.   For example,  it is possible  to calculate
the self-energy  in a self-consistent  manner (parallel with  the charge
density) \cite{MCP+05}.

Introducing  the  antihermitian  part  of the  Green's  function  matrix
(\ref{gfmatk})   ${\Im{G}}_{ij}=\frac{1}{2}  (G_{ij}-G_{ji}^\star)$  and
taking  into  account  translational  symmetry, the hermitian  part  of  the
optical conductivity (\ref{kubo_GF}) is expressed as:
\begin{eqnarray}     \sigma_{\lambda\lambda'}^{(1)}&=&\frac{1}{\pi\omega}
\int  d^3  k\!   \int_{\epsilon_{\rm F}-\hbar\omega}^{\epsilon_{\rm  F}}
\!\!\!\!    d\epsilon  \nonumber   \\  &\times&\sum_{ij}   {\mathcal  J}
^\lambda_{ij}({\bf   k},\epsilon)  {\mathcal   J}^{\lambda'}_{ji}  ({\bf
k},\epsilon+\hbar\omega)
\label{kubolmto}
\end{eqnarray} with 
\begin{eqnarray} {\mathcal  J} ^{\lambda}_{ij}({\bf k},\epsilon)= \sum_n
{\Im          G}_{in}({\bf          k},\epsilon)\langle          n_{{\bf
k}}|\hat{J}^{\,\lambda}|j_{{\bf k}}\rangle\,.
\label{currentdensmat}
\end{eqnarray}    Actually the  possibility to split the 
${\mathcal J}^{\lambda}_{ij}({\bf    k},\epsilon)$   matrix    elements
into   the energy-dependent and -independent parts makes the calculation
of optical conductivity rather fast.

The calculational  procedure  is  built up as  follows. 
The energy-dependent on-site self-energy   $\Sigma(\epsilon)$
 is   obtained   from   the self-consistent  KKR-GF  scheme \cite{MCP+05}.  
For  a  given  self-energy  the  Dyson
equation (\ref{gfmatk}) is used to obtain the effective Green's function
in the basis set of the LMTO method. The  antihermitian part ${\Im G}_{ij}$
of  the effective  Green's  function  is used  to  calculate the  matrix
elements       of       the       current-density      operator       in
Eq.\,(\ref{currentdensmat}).  The  latter  allows  one to  evaluate  the
optical conductivity given by Eq.\,(\ref{kubolmto}). Spin-orbit coupling (SOC) which is together with exchange splitting the actual source of MOKE is taken into account via the
second-variation technique.

\begin{figure}
\includegraphics[width=7.5cm,angle=0]{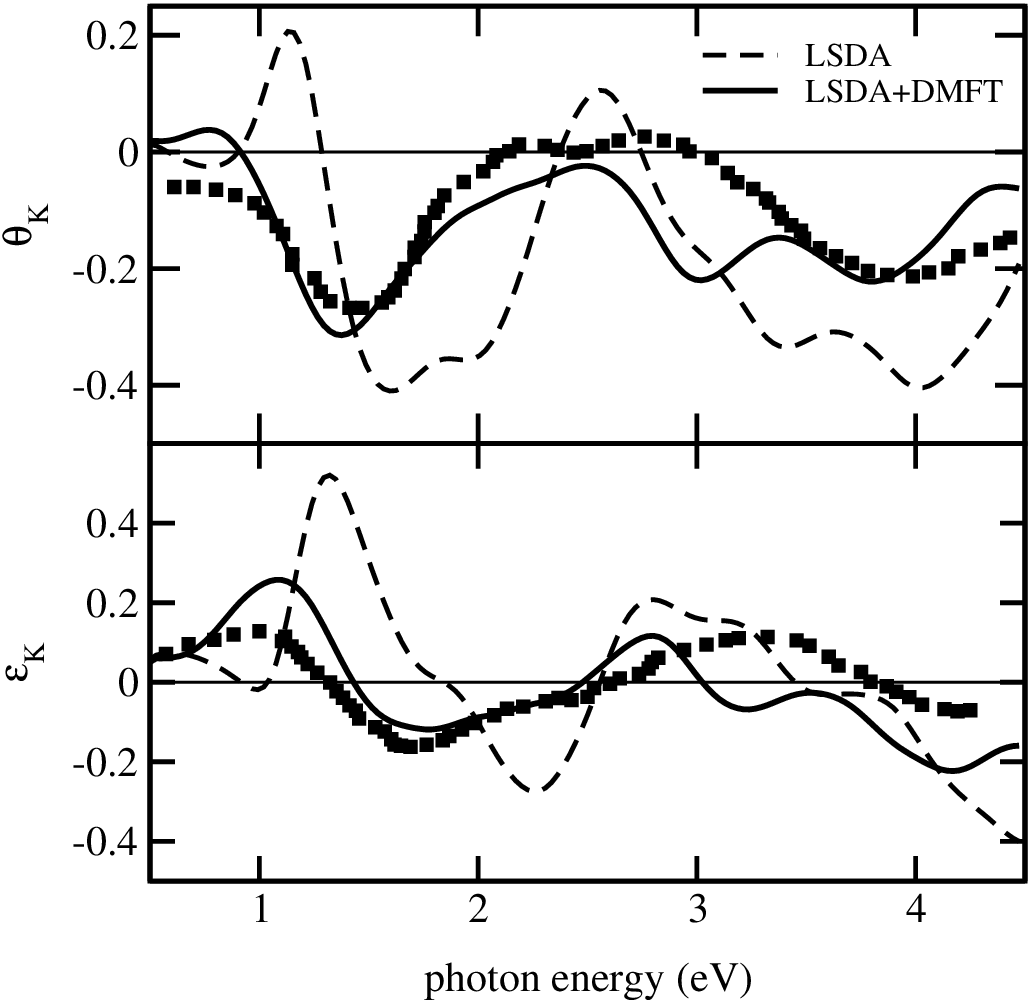}
\caption{\label{kerr_spectr}  MO spectra for  NiMnSb. Upper  panel: Kerr
rotation  angle;  lower  panel:   Kerr  ellipticity.  Solid  line:  LSDA
calculations;  broken line: LSDA+DMFT  calculations.  The  square points
represent the experimental results given in Refs.\,\cite{EBJ+83,Eng83}.}
\end{figure}

The comparsion between the MO  spectra of NiMnSb calculated within LSDA,
LSDA+DMFT  and    the     experimental    results    is    shown    in
Fig.~(\ref{kerr_spectr}).
 The  obvious conclusion is that an  account of
local correlations  is essentially  important to describe  correctly the
positions as  well as the magnitudes  of both low-  and high-energy Kerr
rotation peaks (situated at 1.4~eV and 4~eV).

It is sufficient  to consider in  details only the real part  of the
MOKE spectrum, i.e. the rotation, as the  Kerr ellipticity is a related quantity.

\begin{figure}
\includegraphics[width=7.5cm,]{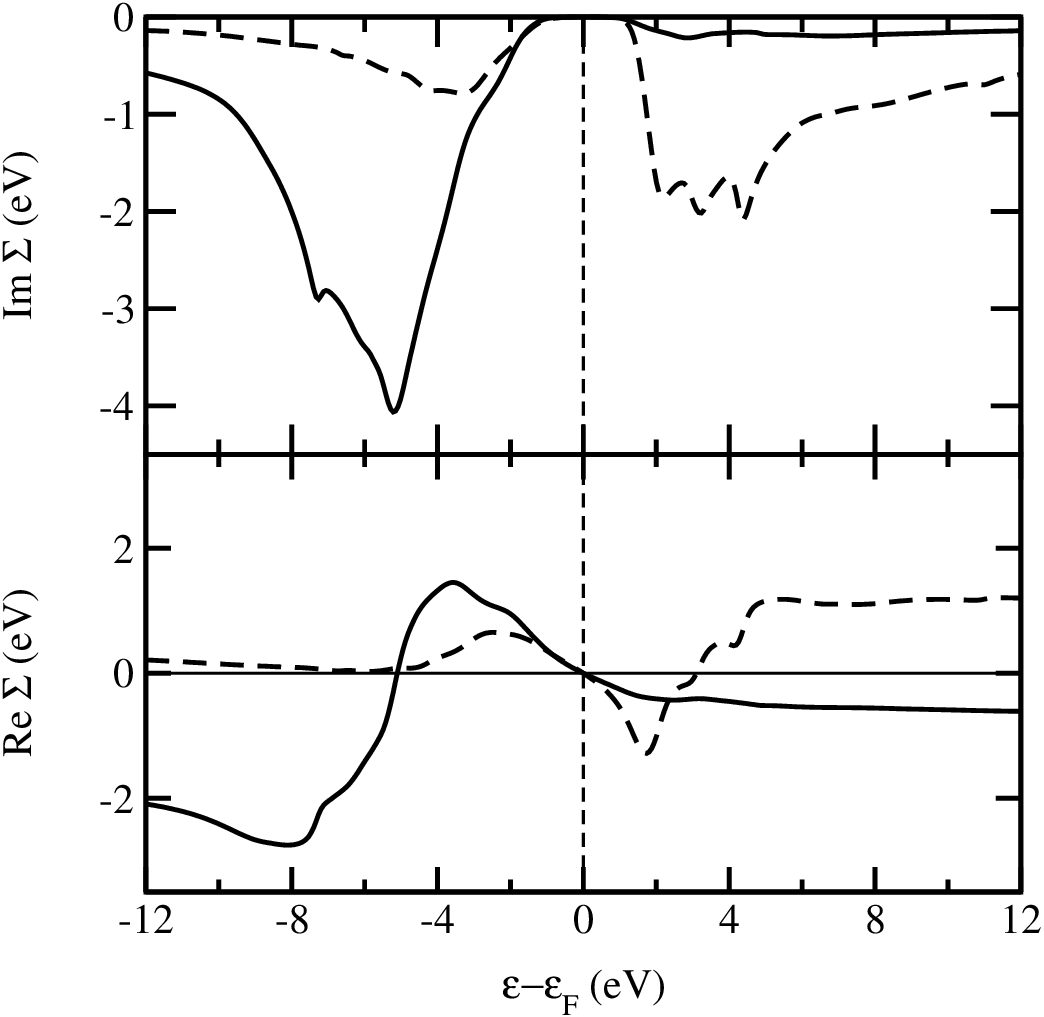}
\caption{\label{SE_Mn}   Spin-resolved  dynamical  self-energy   for  Mn
obtained from SPR-KKR calculations  using SPTF solver. Lower panel: real
component;  upper panel:  imaginary  component. Solid  and broken  lines
correspond to majority and minority spin-components respectively. }
\end{figure}

The Mn $d$-shell indeed experiences noticable dynamical correlations as depicted by
the self-energy plot shown  in Fig.~(\ref{SE_Mn}) with the amplitudes of
the  imaginary  component up  to  4~eV.  For the Ni shell it is  not so  important the account  of
correlations as it  is almost fully  occupied. The
effective   Coulomb  interaction   is  parametrised   by   $U=3$~eV  and
$J=0.9$~eV. Numerical tests show that approximately the same results are
obtained within the range of $U=3\pm$0.5~eV.

\begin{figure}
\includegraphics[width=7.5cm,angle=0]{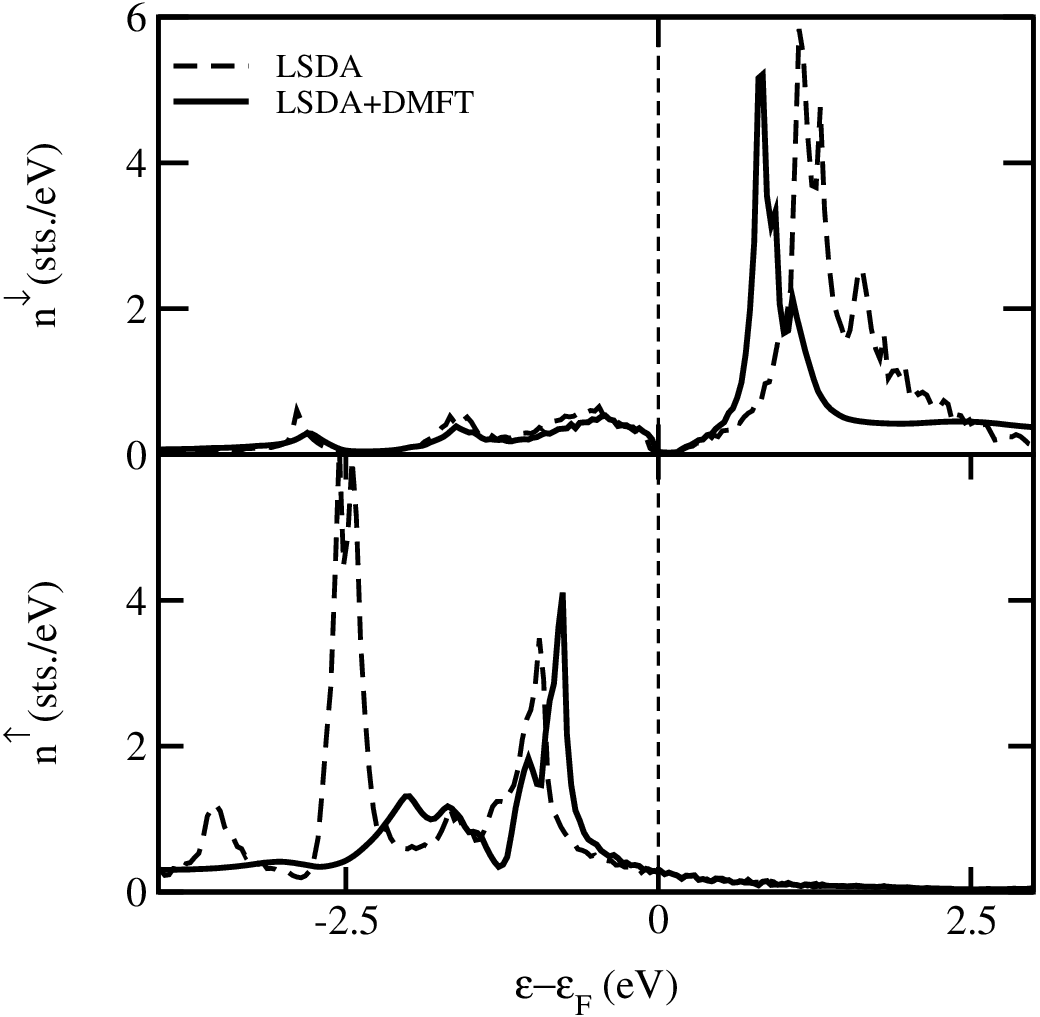}
\caption{\label{DOS_Mn}  Spin-resolved DOS of the Mn $d$-shell.   Solid line:
LSDA calculations; broken line: LSDA+DMFT calculations.}
\end{figure}

Considering   Eqs.\,(\ref{kubolmto}) and (\ref{currentdensmat})   we   can
straightforwardly analyse  the modifications in the MO  spectrum. As the
matrix   elements   $\langle   n_{{\bf   k}}|\hat{J}^{\,\lambda}|j_{{\bf
k}}\rangle$ are not influenced by  DMFT, the two possible sources of the
influence  are  the  Green's  function  matrix and  the  change  in  the
occupation  numbers.  However, in the present  calculations  the number  of
occupied  and unoccupied  states  for each  spin-projection is  strictly
conserved.  This   occurs  due  to  the  special   construction  of  the
self-energy  represented by  the  so-called "particle-particle"  channel
\cite{LK98KL99KL02}. The latter includes the channels of perturbation theory
starting  from the second-order  and that is why pocessing  the dominant
contribution in  the dynamical correlations for  the $d$-electron shell.
This  leads in  particular  to the  conclusion  that local  correlations
modify  only  the  interband  part  of the  optical  conductivity. The
intraband contribution which is determined by the the Green's
function matrix elements at the Fermi level remains unchanged. Thus,  one can  dominantly  relate the  changes  in the Kerr  effect to  the
modifications  in  the interband  contribution  of optical  conductivity
caused  by the  renormalisation of  the one-particle  spectrum  which is
shown in Fig.\,(\ref{DOS_Mn}).
The false peak inserted by DMFT at 3~eV could be attributed first of all
to the lack  of the vertex corrections in  our calculational scheme.  In
order to  describe the  optical transitions correctly,  the two-particle
Green's  function  has  to  be   used.  That is because the  expression  
(\ref{kubo_GF}) formulated in terms of the one-particle Green's function
is an  approximation which needs  to be fullfilled with  the appropriate
vertex  corrections   \cite{Mah90}.   However,  the   account  of  local
correlations   makes  the  implementation   of  vertex   corrections  in
the computational   scheme   rather  complicated   and   needs  a   separate
investigation. Another aspect which might be  interesting to investigate is the account
of nonlocal correlations. In conclusion, it has been demonstrated that an improved description of
correlation  effects on the basis of the LSDA+DMFT scheme leads to a
substantially improved agreement of the theoretical and experimental MO
spectra.

This work was funded by the German BMBF (Bundesministerium f\"ur Bildung
und Forschung) under Contract No. FKZ 05 KS1WMB/1.

\end{document}